\documentclass[prd,onecolumn,nofootinbib,preprintnumbers,floatfix]{revtex4}

\usepackage[plainpages=false, colorlinks=true, anchorcolor=blue, linkcolor=blue, citecolor=blue, bookmarks=false]{hyperref}
\usepackage[utf8]{inputenc}
\usepackage{amsfonts,amsmath,amssymb}
\usepackage{graphicx}
\usepackage{color}
\usepackage{natbib}
\usepackage{enumitem}
\usepackage{subcaption}
\usepackage{comment}
\usepackage{caption}
\usepackage{lipsum}
\newcommand{\rthis}[1]{\textcolor{black}{#1}}

\begin{document}
\title{Comparison of $\Lambda$CDM and $R_h = ct$ with updated galaxy cluster $f_{gas}$ measurements using Bayesian inference}
\author{Kunj \surname{Panchal}}\altaffiliation{E-mail: ep21btech11017@iith.ac.in}

\author{Shantanu  \surname{Desai}}  
\altaffiliation{E-mail: shntn05@gmail.com}

 \newcommand{\apjl}{Astrophys. J. Lett.}
 \newcommand{\apjs}{Astrophys. J. Suppl. Ser.}
 \newcommand{\aap}{Astron. \& Astrophys.}
 \newcommand{\aj}{Astron. J.}
 \newcommand{\araa}{Ann. Rev. Astron. Astrophys. } 
 \newcommand{\aapr}{Astron. \&  Astrophys. Review} 
 \newcommand{\mnras}{Mon. Not. R. Astron. Soc.}
 \newcommand{\apss} {Astrophys. and Space Science}
 \newcommand{\jcap}{JCAP}
 \newcommand{\pasj}{PASJ}
 \newcommand{\na}{New Astronomy}
 \newcommand{\nar}{New Astron. Reviews}
 \newcommand{\pasa}{Pub. Astro. Soc. Aust.}
 \newcommand{\physrep}{Physics Reports}

\begin{abstract}
 We use updated gas mass fraction measurements of 44 massive dynamically relaxed galaxy clusters collated in ~\cite{Mantz22}  to distinguish between the standard $\Lambda$CDM model and $R_h=ct$ universe. For this purpose, we use Bayesian model selection to compare the efficacy of both these cosmological models given the data. The gas mass fraction is modeled using both cosmology-dependent terms and also astrophysical parameters, which account for the variation with cluster mass and 
 redshift. We used two different prior choices for some of the astrophysical parameters. We find a  Bayes factors of 50 and 5 for $\Lambda$CDM  as compared to $R_h=ct$ for these two prior choices. This implies that   $\Lambda$CDM is  favored  compared to $R_h=ct$ with significance ranging from substantial to very strong.
     \end{abstract}  
 \affiliation{Department of Physics, Indian Institute of Technology, Hyderabad, Telangana-502284, India}
 \maketitle
\section{Introduction}
The standard hot Big-Bang model of cosmology described by  a flat $\Lambda$CDM universe, with 70\% dark energy and 25\% cold (non-baryonic) dark matter and 5\% baryons~\cite{Ratra03} agrees with cosmological data at large scales~\cite{Planck2018}. There are however a few data-driven problems with the standard $\Lambda$CDM paradigm, such as the Hubble constant tension between local and high redshift measurements~\cite{Lahav,DiValentino,Verde,Bethapudi,Smoot}, $\sigma_8$ tension between  galaxy clusters and CMB~\cite{Abdalla22}, Lithium-7 problem in Big-Bang nucleosynthesis~\cite{Fields}, etc. An up to date review of all problems and anomalies in $\Lambda$CDM  can be found in ~\cite{Periv,Abdalla22,Peebles22}.
Therefore a large number of alternate models have been proposed to account for some of these anomalies~\citep{alternatives,Banik,Abdalla22}.

One such model is the $R_h=ct$ universe model, proposed by  Melia~\cite{Melia07,Shevchuk,Melia2012}. Here, the size of the Hubble sphere given by $R_h(t)=ct$ is the same at all  times in contrast to the case of the $\Lambda$CDM model, where this coincidence is true only at the current epoch, i.e.  $R_h(t_0)=ct_0$. This model has $H(z)=H_0(1+z)$.
As a consequence of this,  the rate of expansion ($\dot{a}$) is constant; while the  pressure and energy density satisfy an equation of state given by $p=-\frac{\rho}{3}$, where $\rho= \rho_{DE}+\rho_{M}+\rho_r$ and $p=p_{DE}+p_{M}+p_r$~\cite{Melia15}.
In this model,  the scale factor  $a(t) \propto t$~\cite{Melia2016}.  This model has been tested with a whole slew of cosmological observations by Melia and collaborators; such as cosmic chronometers~\cite{Meliamaier,Meliachrono}, quasar core angular size measurements~\cite{Meliaquasar}, quasar X-ray and UV fluxes~\cite{MeliaUV}, Type 1a SN~\cite{MeliaSN}, strong lensing~\cite{MeliaSL,Melialensing23}, BAO~\cite{Melia23,MeliaBAO22}, FRBs~\cite{MeliaFRB} and  found to be in better agreement compared to $\Lambda$CDM model.  The $R_h=ct$ model can also explain the preponderance of massive galaxies with $z>10$ observed by JWST~\cite{MeliaJWST}.
However, other researchers have reached opposite conclusions and have argued that this model is  either inconsistent  with observations or is not favored  compared to  $\Lambda$CDM~\cite{Shafer,Seikel_2012_rhct,LewisBBN,Haridasu1,Lin,Hu2018,Tu2019,Fuji,Haveesh}. However, many of the above works have used BAO and Alcock-Pacynzski measurements to rule out $R_h=ct$ models. The  BAO measurements are scaled by the size of the sound horizon at the drag epoch, $r_s$, whose calculation implicitly assumes the $\Lambda$CDM model~\cite{Meliamaier,Meliachrono}. The latest model-independent tests done using BAO without assuming a value of $r_s$ or $H_0$, based on Lyman-$\alpha$ measurements favor $R_h=ct$ over $\Lambda$CDM~\cite{Melia23}.
Therefore, it is imperative  to do the comparison with cosmological observables, which are independent of the underlying theoretical model. In this work, we revisit the question of comparison of $\Lambda$CDM and $R_h=ct$  using another model-independent probe, viz. gas mass fraction measurements in galaxy clusters. For this purpose, we use updated measurements  collated in ~\cite{Mantz22} (MMA22 hereafter),  to follow-up on similar test previously carried out in ~\cite{Meliafgas} (M16, hereafter) using three different datasets.

 The outline of this manuscript is as follows. We first review the analysis carried out in M16 in Sect.~\ref{sec:m16}.  A very brief primer on Bayesian model comparison can be found in Sect.~\ref{sec:bayesian}. We discuss the latest $f_{gas}$ data compiled in MMA22 in Sect.~\ref{sec:latestdata}.
 Our analysis and results are summarized in Sect.~\ref{sec:results}. We conclude in Sect.~\ref{sec:conclusions}.

 \section{Recap of M16}
 \label{sec:m16}
 Galaxy clusters are the most massive virialized objects in the universe and provide wonderful laboratories for studying a wide range of topics from galaxy evolution to cosmology~\cite{Vikhlininrev,allen11,Kravtsov2012} to fundamental Physics~\cite{Desai,Bohringer16, Boraalpha, BoraDesaiCDDR, BoraDM,Holandac}.  About 10-15\% 
 of the mass content comprises of diffuse hot gas
 in the  intracluster  medium (ICM). The gas mass fraction defined as the ratio of gas mass to the total mass ($f_{gas} \equiv M_{gas}/M_{tot}$) for dynamically relaxed clusters has been used as a cosmological probe in a large number of works
~\cite{allen02,allen04,allen08,Ettori09,Mantz14}. 

The key assumption in the above works  is that $f_{gas}$ at the measured overdensity is constant and independent of redshift~\cite{white93,Sasaki96,Pen,lima03}. This {\it ansatz} has been tested using a large number of simulations over the past two decades. In ~\cite{Kravtsov05}, it was shown that the cumulative baryon fraction is close to the universal values of $\Omega_b/\Omega_{DM}$ for $r \geq r_{2500}$. ~\citet{Battaglia} found a constant gas fraction at $r_{500}$ using simulations having multiple prescriptions for radiative gas dynamics, star formation, and AGN feedback. Then, ~\citet{planelles13} studied the baryon fraction within $R_{2500}$, $R_{500}$, and $R_{200}$ by using three different suites of  simulations  (with each simulation covering a different physical process) of  galaxy clusters in the redshift range $0 \leqslant z \leqslant 1$. They found that the baryon 
baryon fraction is the same for all the three physical processes and has negligible redshift evolution for $R_{200}$, $R_{500}$, and $R_{2500}$.  Most recently, ~\citet{Ange23} studied the redshift  evolution of gas mass fraction  using the {\it Magneticum}  simulations and quantified  the evolution using the gas depletion parameter $\gamma (z)  \equiv f_{gas}/(\Omega_{bar}/\Omega_{tot})$. They found that $\gamma(z)$ shows a lot of scatter between $(0.5-2) R_{500}$ for less massive haloes, while for larger radii $> 2 R_{500}$, it  is independent of mass and redshift~\cite{Ange23}.  A large number of works have also done model-independent tests of the redshift evolution of the gas depletion factor ($\gamma (z)$) using observations of X-Ray and Sunyaev-Zeldovich  (SZ) selected galaxy clusters  at both $r_{500}$ and $r_{2500}$~\cite{holanda1706,holanda1711,zheng19,BoraDesaifgas,HolandaBoraDesai}. We report a few highlights from these works. For measurements at $r_{2500}$,  no evolution of the gas depletion factor with redshift was found in \cite{holanda1706,holanda1711}. However, a mild decrease in the redshift evolution of gas depletion factor was found in some other works~\cite{zheng19,HolandaBoraDesai}. The first work did not use direct $f_{gas}$ from X-ray observations, but used an empirical relation to obtain $f_{gas}$. The second work used strong gravitational lensing systems as distance anchors. A model-independent test of the gas depletion factor using two different SZ selected samples (SPT-SZ~\cite{Chiu18} and Planck-ESZ~\cite{PlanckESZ}) based on direct X-ray measurements at $r_{500}$ found opposite trends for the two samples and also a strong dependence on the halo mass~\cite{BoraDesaifgas}.

The model used for  $f_{gas}$  in M16   can be written in terms of the reference $\Lambda$CDM model (denoted by the superscript {\it ref}) and an alternate cosmological model as follows:
\begin{equation}
    f_{gas}^{model} = K \left[ \frac{H(z)d_A(z)}{[H(z)d_A(z)]^{ref}} \right]^\eta \left[\frac{d_A^{ref}(z)}{d_A(z)} \right]^{3/2}.
    \label{eq:fgas}
\end{equation}
The terms in the numerator and denominator in the first and second parenthesis, respectively refer to the cosmological model, which is  being tested. The reference model used is the flat  $\Lambda$CDM  with  $\Omega_m=0.3$, $\Omega_{\Lambda}=0.7$, and $H_0=70$ km/sec/Mpc.
In Eq.~\ref{eq:fgas}, $d_A(z)$ is the angular-diameter distance, $H(z)$ is the Hubble parameter, and $K$ parameterizes the instrumental calibration uncertainty. For flat $\Lambda$CDM, $H(z)= H_0\sqrt{\Omega_m(1+z)^3+ 1-\Omega_m}$, For $R_h=ct$, $H(z)=H_0(1+z)$. For $R_h=ct$ universe, $d_A$ is given by~\cite{Meliafgas}:
\begin{equation}
d_A = \frac{c}{H_0 (1+z)} \ln (1+z) 
\label{eq:darh=ct}
\end{equation}
For flat $\Lambda$CDM,  $d_A$ is given by~\cite{Hogg99}:
\begin{equation}
    d_{A} = \frac{c}{H_0} \left(\frac{1}{1+z}\right) \int_{0}^{z}\frac{dz^{'}}{\sqrt{\Omega_m{(1+z^{'})^3}+(1 - \Omega_m)}}
   \label{eq:ADD}
\end{equation}

M16 applied this formalism to three different $f_{gas}$ datasets~\cite{Laroque06,allen08,Ettori09}. The first two of these datasets  consist of gas fraction measurements at $R_{2500}$~\cite{Laroque06,allen08}, while the third used measurements at $R_{500}$~\cite{Ettori09}. All the three datasets were separately fit to Eq~\ref{eq:fgas} for both the models using $\chi^2$ minimization. The $R_h=ct$ model has no free parameters, while $\Lambda$CDM has 1-2 free parameters (depending on whether one assumes a flat model or not).
Both cosmologies were consistent with a constant $f_{gas}$ ratio. The reduced $\chi^2$ for $\Lambda$CDM and $R_h=ct$ were less than one for all the  three samples. The efficacy of these models was evaluated using information theoretical techniques such as AIC, BIC,  and KIC~\cite{Liddle_2007,Krishak}. For all the three datasets the results from BIC favored $R_h=ct$ with a likelihood of $\sim 95\%$ as compared to 5\% for $\Lambda$CDM. Therefore, M16 argued that the gas mass fraction data favored $R_h=ct$ over $\Lambda$CDM. However, we note that these information theoretical criteria assume that the posterior distributions are Gaussian or close-to Gaussian, which may not always be satisfied~\cite{Liddle_2007}. Therefore, we shall use Bayesian model comparison techniques for model selection.

\section{Bayesian Model Comparison}
\label{sec:bayesian}
Before, we discuss the dataset used for our analysis, we provide a very  brief prelude to   Bayesian model comparison used to compare the relative efficacy of $\Lambda$CDM and $R_h=ct$. More details can be found  in recent reviews~\cite{Trotta,Weller,Sanjib,Krishak} or our most recent works~\cite{Desai23,Srivastava23,Pasumarti23}.

To evaluate the significance of one  model ($M_2$) as compared to another one ($M_1$), we calculate the Bayes factor ($B_{21}$) given by:
\begin{equation}
B_{21}=    \frac{\int P(D|M_2, \theta_2)P(\theta_2|M_2) \, d\theta_2}{\int P(D|M_1, \theta_1)P(\theta_1|M_1) \, d\theta_1} ,  \label{eq:BF}
\end{equation}
where $P(D|M_2,\theta_2)$ is the likelihood for the model $M_2$ given the data $D$ and $P(\theta_2|M_2)$ denotes the prior on the parameter vector $\theta_2$ of the model $M_2$.  The denominator denotes the same for model $M_1$. The full expression in the numerator and denominator is referred to as Bayesian evidence for each of the two models. 
If $B_{21}$ is greater than one, then $M_2$ is  preferred over $M_1$ and vice-versa. The significance can be qualitatively assessed depending on the value of the Bayes factor based on the Jeffreys' scale~\cite{Trotta}. Note that unlike the frequentist or information theory based model comparison techniques, Bayesian evidence does not use the best-fit value for the likelihood or posterior, and does not make any assumptions about the posteriors~\cite{Sanjib}. Another advantage of Bayesian model comparison  is that it does not penalize parameters which are unconstrained by the data unlike information theory techniques~\cite{Liddle_2007}.
Therefore, it is the most robust among the different model comparison techniques.

\section{Dataset used for analysis}
 \label{sec:latestdata}
We summarize the dataset used in MMA22 for this analysis, where more details can be found. MMA22 considered 44 clusters with Chandra X-ray observations in the redshift range $0.018 \leq z \leq 1.16$. These clusters are dynamically the most relaxed with their intra-cluster medium temperatures $\geq$ 5 keV. This minimizes any systematic uncertainties related to departures from hydrostatic equilibrium. The gas measurements were obtained in a shell within $(0.8-1.2) r_{2500}$ of each cluster.
40 of these clusters  were used for previous cosmological studies with $f_{gas}$~\cite{Mantz14}, and MMA22 used deeper   observations for the same clusters. In addition, four other clusters, viz. Perseus Cluster,
RCS J1447+0828, SPT J0615-5746, and SPT J2215-3537 were added to the sample. The total exposure time for these clusters was equal to  4.9 Ms. A tabulated summary of the $f_{gas}$ measurements for these clusters along with their associated uncertainties can be found in Table 1 of MMA22. Similar to previous works, all the $f_{gas}$ measurements were tabulated assuming a reference cosmology of $\Omega_M=0.3, \Omega_{\Lambda}=0.7$ and  $H_0=70$km/sec/Mpc. The X-ray based masses for these clusters have also been compared to weak lensing masses, with their ratio given by $0.96 \pm 0.12$~\cite{Applegate16}, where the error is  the quadrature sum of statistical and systematic errors. Therefore, there is no hydrostatic bias for this cluster sample.

\section{Analysis and Results}
\label{sec:results}
 
  \subsection{Model}
    Similar to MMA22, our model for the  gas mass fraction also includes various astrophysical nuisance parameters in addition to the dependence on cosmology. We use the same  parameterization  as  in MMA22 and is described as follows:
    \begin{equation}
        f_{gas} (z,M_{2500}) = \gamma(z,M_{2500}) \frac{\Omega_b}{\Omega_m},
        \label{eqn:5}
    \end{equation}
where $\Omega_b$ and $\Omega_m$ are the matter densities. \rthis{For their  values we  use measurements from  Planck 2020 cosmology, viz. $\Omega_m=0.315 \pm 0.0011$ and $\Omega_b=0.05 \pm 0.008$~\cite{Planck2018}, which are obtained from a combination of TT, EE and TE angular power spectra (cf. Table 1 of ~\cite{Planck2018}).}  
This ratio is a measure of the baryonic to dark matter density (or the cosmic baryon fraction), which we assume (similar to M16) is independent of redshift for our cluster sample. Note that since this  ratio  will appear in the expression for both $\Lambda$CDM as well as $Rh=ct$, hence we keep these fixed, even though $\Omega_m$ and $\Omega_b$ are prima-facie free parameters within $\Lambda$CDM. 
It might also seem counter-intuitive to use these same values for $R_h=ct$, since it is sometimes stated  that this model might not contain dark matter~\cite{Lewis12}. However, as emphasized by Melia and collaborators in multiple works~\cite{Melia15,Meliafgas,Reddy21,Fatuzzo}, the $R_h=ct$ model also contains baryonic and dark matter, in addition to dark energy and radiation. The value of $\Omega_M$ has been estimated to be  $\approx 0.27$~\cite{Melia15}. Similarly it was shown that  $\Omega_b$ must satisfy the constraint $0.026 \lesssim \Omega_b \lesssim 0.037$ to reconcile with constraints on reionization~\cite{Fatuzzo}. Furthermore, M16 has stressed  that because of Birkhoff's theorem, if $f_{gas}$ (which is $\propto \Omega_b/\Omega_m$) is constant in $\Lambda$CDM, it should be the same in all cosmological models, including $R_h=ct$. Therefore, it is safe to assume that this ratio is constant for $z \lesssim 2$ as emphasized in M16. Therefore, we can use the Planck 2020 value for the ratio of $\Omega_b$ to $\Omega_m$  in Eq.~\ref{eqn:5} for both the models.

Finally, $\gamma(z,M_{2500})$ in Eq.~\ref{eqn:5} is the gas depletion factor and  is parameterized according to:
    \begin{equation}
        \gamma(z,M_{2500}) = \gamma_0 (1 + \gamma_1 z) \left( \frac{M_{2500}}{3 \times 10^{14} M_{\odot} }\right)^{\alpha}.
    \end{equation}
Therefore,  the gas depletion factor has an  explicit dependence on the halo mass. Such a dependence  of the gas  fraction on the halo mass has also been found in model-independent estimates of the  gas depletion factor in SPT-SZ and Planck ESZ data~\cite{BoraDesaifgas}.
Finally,  we incorporate the cosmology dependence of $f_{gas}$ as follows: 
   \begin{equation}
       f_{gas}^{ref} = K(z) {A(z)}^{\eta_f} \left[ \frac{d^{ref}(z)}{d(z)} \right ]^{3/2} f_{gas}(z,M_{2500}),
       \label{eq:fgasref}
   \end{equation}
where    $f_{gas}^{ref}$ is the gas mass fraction for the reference cosmology (flat $\Lambda$CDM with $\Omega_M=0.3$ and $\Omega_{\Lambda}=0.7$) and $f_{gas}$  is given in Eq.~\ref{eqn:5}; $A(z)$ represents corrections in the measurement aperture due to the reference cosmological model and  is parameterized in the same way as MMA22:
    \begin{equation}
         A(z) = \frac{\theta^{ref}_{2500}}{\theta_{2500}} = \frac{\left[H(z) d(z) \right ]}{\left [H(z) d(z) \right ]^{ref}}
    \end{equation}
Finally,  $K(z)$ represents a redshift dependent bias in the gas fraction measurements due to a bias in the total mass estimates and can be written as:
    \begin{equation}
        K(z) = K_0 (1 + K_1 z) .
    \end{equation}
\subsection{Likelihood and Priors}   
   To calculate the Bayesian evidence for both the models, we use  the following Gaussian likelihood: 
        \begin{equation}
             -2 \ln{\mathcal{L}} = \sum_{i} \ln{2\pi {\sigma_i}^2} + \sum_{i} {\left( \frac{f_{obs,i} - f_{gas}^{model}}{\sigma_i}\right)}^2, 
        \end{equation}
  where       
   $f_{gas,i}$ is the  observed gas mass fraction of galaxy cluster,
   $f_{gas}^{model}$ is the cosmology-dependent model for  $f_{gas}$ outlined in Eq.~\ref{eq:fgasref}, and 
   $\sigma_i$ is the observed error in $f_{obs,i}$. While evaluating $f_{gas}^{model}$ for $\Lambda$CDM, we use $D_A$ given in Eq.~\ref{eq:ADD} and for $R_h=ct$ we use $D_A$ given in Eq.~\ref{eq:darh=ct}.
    The priors used for all the free parameters  in Eq.~\ref{eq:fgasref} are  the same as in MMA22 and are summarized in  Table~\ref{table:lcdm}. Furthermore, we also redid the analysis with Gaussian priors on $K_0$ and $\gamma_0$ similar to a recent work, which has used the same data for determination of Hubble constant~\cite{Gonzalez24} (See also ~\cite{Barbosa24} which used Gaussian prior on $K_0$ while doing another cosmological tests using gas mass fraction data). We present our results for both these choice of priors.
     We note that $R_h=ct$ has no cosmology related free parameters, and the only free parameters in this model   are the astrophysical nuisance parameters,  which are also common to $\Lambda$CDM.  For  the flat $\Lambda$CDM  model, we have an additional free parameter, namely  $\Omega_m$, used in calculation of angular diameter distance.

   \begin{table}[h!]
       \begin{center}
               \renewcommand{\arraystretch}{1.5}
               \begin{tabular}{|c|c|}
               \hline
               \textbf{Parameter} & \textbf{Prior} \\
               \hline
               $K_0$ & $\mathcal{U}$ (0,2) \\
                        & $\mathcal{N}$  (0.93,0.11) (*)\\
               $K_1$ & $\mathcal{U}$ (-0.05,0.05) \\
                     & $\mathcal{U}$ (-0.6,0.6) (*)\\
               $\gamma_0$ & $\mathcal{U}$ (0.71,0.87) \\
                          & $\mathcal{N}$ (0.79,0.07) (*) \\
               $\gamma_1$ & $\mathcal{U}$ (-0.05,0.05) \\
                          & $\mathcal{U}$ (-0.6,0.6)(*) \\
               $\alpha$ & $\mathcal{U}$ (-1,1) \\
               $\eta_m$ & $\mathcal{N}$ (1.065,0.016) \\
               $\eta_f$ & $\mathcal{N}$ (0.390,0.024) \\
               $\Omega_m$& $\mathcal{U}$ (0,1) \\
               \hline
               \end{tabular}
        \end{center} 
         \caption{Priors used for the free parameters for $\Lambda$CDM and $R_h=ct$. Except for $\Omega_m$, which exists only in the calculation of $D_A$ in $\Lambda$CDM, the same priors were used for both the models and are the same as in MMA22. We did our analyses using two different prior choices for $\gamma_0$, $\gamma_1$, $K_0$ and $K_1$. The second set of priors is denoted by the asterisk.}
          \label{table:lcdm}
    \end{table}

To  calculate the Bayesian evidence for both the models, we have the used {\tt Dynesty}~\cite{Speagle} package, which is based on the Nested sampling algorithm.~\cite{NS}. We now present our results.
\begin{itemize}
\item \textbf{Uniform priors on $K_0$ and $\gamma_0$:}
When we use  the same uniform priors on $K_0$ and $\gamma_0$ as in MMA22, the Bayes factor for the flat  $\Lambda$CDM  model compared to $R_h=ct$ model is equal to 50. According to Jeffreys' scale~\cite{Weller} this constitutes ``very strong'' evidence for flat  $\Lambda$CDM over $R_h=ct$. 
\item \textbf{Normal priors on $K_0$ and $\gamma_0$:}
Once again, we  calculate the Bayes factors using normal priors on $K_0 \in \mathcal{N}$(0.93,0.11) and $\gamma_0 \in \mathcal{N}$ (0.79,0.07) similar to \cite{Gonzalez24}. Since we do not get closed contours for $K_1$ and $\gamma_1$,
using these normal priors, we use a broader range of priors on
$K_1$ and $\gamma_1$ given by $K_1 \in \mathcal{U}(-0.6,0.6)$ and $\gamma_1 \in \mathcal{U}$(-0.6,0.6). With these new priors we find Bayes factor of around 5, which corresponds to substantial evidence in favor of the flat $\Lambda$CDM model.
\end{itemize}

Therefore, even though earlier analysis in M16 favored $R_h=ct$ over $\Lambda$CDM, updated $f_{gas}$ measurements at $R_{2500}$ collated in MMA22 seem to  favor $\Lambda$CDM with very strong or substantial evidence depending on whether uniform or normal priors are used on $K_0$ and $\gamma_0$. Of course,  $\Lambda$CDM, is still not decisively favored, for which the Bayes factor needs  to be greater than 100. In the Appendix, we also apply this procedure to one of the other datasets which used $f_{gas}$ measurements at $R_{2500}$~\cite{allen08}.

\section{Conclusions}
\label{sec:conclusions}
A test for the  constancy of  galaxy cluster gas mass fraction measurements as a function of redshift  was implemented in M16  for two different cosmological models,  $\Lambda$CDM and $R_h=ct$ using three different datasets. They found that both the models are consistent with a constant gas mass fraction. However, since the $R_h=ct$ model has no free parameters, M16 asserted  using BIC based information theory test that $R_h=ct$ is  favored over $\Lambda$CDM with a relative likelihood percentage  of 95:5.

Recently, updated gas mass fraction measurements at $R_{2500}$ of one of the samples tested in M16,  were presented in MMA22. This sample consists of 44 dynamically relaxed very massive clusters. In this work,  we do a similar comparison of the above two  cosmological models with these  updated gas mass fraction measurements. For this purpose, we use Bayesian model selection and calculate  the Bayes factor between these two models to ascertain the best model among the two. Furthermore, we model the gas mass fraction using additional astrophysical nuisance parameters, which account for the dependence on cluster mass and redshift, in addition to the cosmology-dependent terms. We did our analyses with two sets of priors. For the first set, we used the same priors as in MMA22. For the second set we used Gaussian priors on $\gamma_0$ and $K_0$, similar to another recent work~\cite{Gonzalez24}, whereas MMA22 had used uniform priors for the same. 
When we use the same priors as MMA22,  we find that the Bayes factor for the flat $\Lambda$CDM compared to  $R_h=ct$ is about 50. According to the Jeffreys' scale, this constitutes very strong evidence for  the flat $\Lambda$CDM over $R_h=ct$. However, when we use Gaussian priors for two astrophysical nuisance parameters $\gamma_0$ and $K_0$, we get a Bayes factor of 5, which corresponds to substantial evidence in favor of $\Lambda$CDM. Therefore, we find that the Bayes factors depends on the priors used. However for both the prior choices we find moderate to strong evidence in favor of  the flat $\Lambda$CDM model over $R_h=ct$.
Therefore, we conclude that contrary to M16 the latest gas mass fraction measurements at $r_{2500}$ support  $\Lambda$CDM over $R_h=ct$.

\begin{acknowledgments}
We are grateful to Adam Mantz for clarification about priors used in MMA22 and   the anonymous  referee for several useful comments and constructive feedback on our manuscript.
\end{acknowledgments}

\bibliography{references}

\appendix
\section{}
\label{Appendix}
We carried out the same analysis done in Sect.~\ref{sec:results} to the data given in \cite{allen08}, which was one of the datasets used in M16 and consists of $f_{gas}$ measurements at $r_{2500}$~\cite{allen08}. This data was obtained by the same team which created the MMA22 dataset.  Note however that this dataset has been superseded by the MMA22 dataset.
The dataset  in ~\cite{allen08} consists of  42 hot dynamically relaxed galaxy clusters with $kT_{2500} > 5$~keV observed using Chandra in the redshift range of $0.05 < z < 1.1$ with a total exposure time of 1.63 Ms. Out of the 42 clusters in \cite{allen08}, 29 are common with the data collated in MMA22. The dataset did not provide values for $M_{2500}$, but listed   $r_{2500}$ measurements for each of the clusters.  We calculated $M_{2500}$ using the $r_{2500}$ values using the following relation:
\begin{equation}
    M_{\Delta} = \frac{4 \pi}{3} \Delta  \rho_c (z)  r^3,    
\end{equation}
with $\Delta = 2500$. Here $\rho_c (z)$ is the critical density at the redshift of the galaxy cluster and corresponds to  and $r$ is the radius  within which the mean overdensity of cluster is $\Delta$ times the critical density. The critical density at a redshift $z$ is given by
\begin{equation}
    \rho_c (z) = \frac{3H^2(z)}{8 \pi G} 
\end{equation}
We use the same reference cosmology as \cite{allen08}, which is a flat $\Lambda$CDM with $\Omega_m$ = 0.3 and $H_0$ = 70 km/s/Mpc as a reference cosmology to carry out the analysis.  We first did our analysis using the same prior as in MMA22 for this data. However, since we could not get closed contours for $K_1$ and $\gamma_1$, we used extended range of  uniform priors for these parameters to $\mathcal{U}$ (-1,1). We obtain a Bayes factor of 1.5 for $\Lambda$CDM over $R_h=ct$. When we used the same Gaussian priors for $K_0$ and $\gamma_0$ as in ~\cite{Gonzalez24}, we get a Bayes factor of 1.2 Therefore with both prior choices, both the models are equally favored based on Bayesian model comparison and one cannot arbitrate between the two models. Therefore, unlike M16, we do not find any evidence that $R_h=ct$ is favored over the flat $\Lambda$CDM for the dataset in ~\cite{allen08}.

\end{document}